\documentclass{sig-alternate-05-2015}
\newcommand{\BfPara}[1]{{\noindent\bf#1.}}
\usepackage{amsmath}
\usepackage{xspace}
\newcommand{\etal}{{\em et al.}\xspace}
\usepackage{subfigure}
\usepackage[utf8]{inputenc} 

\usepackage{graphicx}

\newenvironment{conditions} 														
  {\par\vspace{\abovedisplayskip}\noindent\begin{tabular}{>{$}l<{$} @{${}:{}$} l}} 	
  {\end{tabular}\par\vspace{\belowdisplayskip}}									 	
\usepackage{siunitx} 																
\usepackage{amsmath}		
\usepackage{multirow}
\newcolumntype{L}[1]{>{\raggedright\let\newline\\\arraybackslash\hspace{0pt}}m{#1}}
\newcolumntype{C}[1]{>{\centering\let\newline\\\arraybackslash\hspace{0pt}}m{#1}}
\newcolumntype{R}[1]{>{\raggedleft\let\newline\\\arraybackslash\hspace{0pt}}m{#1}}

\begin{document}

\title{Multimodal Game Bot Detection using User Behavioral Characteristics}

\author{Ah Reum Kang \\ University at Buffalo \and Seong Hoon Jeong \\ Korea University \and Aziz Mohaisen \\ University at Buffalo \and Huy Kang Kim \\ Korea University}

\maketitle 

\begin{abstract} 
As the online service industry has continued to grow, illegal activities in the online world have drastically increased and become more diverse. Most illegal activities occur continuously because cyber assets, such as game items and cyber money in online games, can be monetized into real currency. The aim of this study is to detect game bots in a Massively Multiplayer Online Role Playing Game (MMORPG). We observed the behavioral characteristics of game bots and found that they execute repetitive tasks associated with gold farming and real money trading. We propose a game bot detection methodology based on user behavioral characteristics. The methodology of this paper was applied to real data provided by a major MMORPG company. Detection accuracy rate increased to 96.06\% on the banned account list.
\end{abstract}



\section{Background}
A game bot is an automated program that plays a given game on behalf of a human player. Game bots can earn much more game money and items than human users because the former can play without requiring a break. Game bots also disturb human users because they consistently consume game resources. For instance, game bots defeat all monsters quite rapidly and harvest items, such as farm produce and ore, before human users have an opportunity to harvest them. Accordingly, game bots cause complaints from human users and damage the reputation of the online game service provider. Furthermore, game bots can cause inflation in a game's economy and shorten the game's lifecycle, which defeats the purpose for which game companies develop such games~\cite{bib24}.

Several studies for detecting game bots have been proposed in academia and industry. These studies can be classified into three categories: client-side, network-side, and server-side. Most game companies have adopted client-side detection methods that analyze game bot signatures as the primary measure against game bots. Client-side detection methods use the bot program's name, process information, and memory status. This method is similar to antivirus programs that detect computer viruses~\cite{MohaisenA14}. Client-side detection methods can be readily detoured by game bot developers, in addition to degrading the computer's performance. For this reason, many countermeasures that are based on this approach, such as commercial anti-bot programs, are not currently preferred. 

Network-side detection methods, such as network traffic monitoring or network protocol change analysis, can cause network overload and lag in game play, a significant annoyance in the online gaming experience. To overcome these limitations of the client-side and network-side detection methods, many online game service providers employ server-side detection methods. Server-side detection methods are based on data mining techniques that analyze log data from game servers. Most game servers generate event logs whenever users perform actions such as hunting, harvesting, and chatting. Hence, these in-game logs facilitate data analysis as a possible method for detecting game bots. 

Online game companies analyze user behaviors or packets at the server-side, and then online game service providers can selectively block those game bot users that they want to ban without deploying additional programs on the client-side. For that, most online game service providers prefer server-side detection methods. In addition, some online game companies introduced big data analysis system approaches that make use of data-driven profiling and detection~\cite{bib24}.  Such approaches can analyze over 600 TB of logs generated by game servers and do not cause any side-effects, such as performance degradation or conflict with other programs.

The literature is rich of various works on the problem of game bot detection that is summarized in Table \ref{table1},  which compares various server-side detection schemes classified into six analysis categories: action frequency, social activity, gold farming group, sequence, similarity, and moving path. Each of those techniques, as surveyed in section~\ref{sec:related}, has advantages and disadvantages; none of the techniques look at the multimodality of the features utilized of detection, which is a step we take in this paper.

\BfPara{Contribution} To this end, we collaborated with NCSoft, Inc., one of the largest MMORPG service companies in South Korea, in order to analyze long-term user activity logs and understand discriminative features for high fidelity bot detection. In this paper, we propose a game bot detection framework. Our framework utilizes multimodal users' behavioral characteristic analysis and feature extraction to improve the accuracy of game bot detection. We adopted some features discovered in the prior literature in confirmed in our analysis, as well as some new features discovered in this study. We combine those features in a single framework to achieve better accuracy and enable robust detection. An additional contribution of this work is also the exploration of characteristics of the misclassified users and bots, highlighting plausible explanations that are in line with users and bots features, as well as the game operations.

\section{Related Work} \label{sec:related}

\begin{table*}[htb]
\begin{center}
\caption{Previous research on server-side detection.}
\begin{tabular}{p{4cm}p{7cm}p{5cm}}
\hline
Category & Definition/key papers & Key idea\\ \hline
Action frequency analysis & Detection method based on users' game play pattern analysis~\cite{bib1, bib2, bib3, bib4, bib5} & - Action frequency, type, and time-interval analyses \newline - Idle time analysis\\ \hline
Social activity analysis & Detection method based on users' social interactions analysis~\cite{bib6, bib7, bib8, bib9} & - Party play log analysis \newline - Chatting pattern analysis \newline - Social network analysis\\ \hline
Gold farming group analysis & Detection method based on users' economic activities analysis~\cite{bib10, bib11, bib12,woo2011can} & - Real money trading analysis \newline - Trade network analysis \newline - Connection pattern analysis\\ \hline
Sequence analysis & Detection method based on users' continuous play sequences analysis~\cite{bib13, bib14, bib15} & - Game event sequence analysis \newline - Combat sequence analysis\\ \hline
Similarity analysis & Detection method based on users' behavioral pattern similarity analysis~\cite{bib16, bib24} & - Self-similarity analysis\\ \hline
Moving path analysis & Detection method based on patterns and zones of moving path analysis~\cite{bib17, bib18, bib19, bib20, bib21} & - Coordinate analysis \newline - Zone analysis\\ \hline
\end{tabular}
\label{table1}
\end{center}
\end{table*}

Action frequency analysis uses the fact that the frequencies of particular actions by game bots are much higher than that of human users. To this end, Chen \etal \cite{bib1} studied the dynamics of certain actions performed by users. They showed that idle and active times in a game are representative of users and discriminative of users and bots.
Thawonmas \etal \cite{bib2} utilized the information on action frequencies, types, and intervals in MMORPG log data.
To detect game bots, Park \etal \cite{bib3} selected six game features, namely map changes, counter-turn, rest states, killing time, experience point, and stay in town. 
Chung \etal \cite{bib4} were concerned with various game play styles and classified them into four player types: killers, achievers, explorers, and socializers.
Zhang \etal \cite{bib5} clarified user behaviors based on game playing time. While this approach provides high accuracy, it is limited in several ways. First, they only focus on observations of short time window, thus they are easy to evade. Second, some of such work focuses only on a limited feature space, thus the approach is prone to confusing bots with ``hardcore'' users (users who use the game for long times; who are increasingly becoming a phenomenon in the online gaming communities). 

Social activity analysis uses the characteristics of the social network to differentiate between human users and game bots. Varvello \etal \cite{bib6} proposed a game bot detection method emphasizing on the social connections of players in a social graph. 
Our previous study chose chat logs that reflect user communication patterns and proposed a chatting pattern analysis framework~\cite{bib7}. 
Oh \etal \cite{bib8} used the fact that game bots and human users tend to form respective social networks in contrasting ways and focused on the in-game mentoring network. 
Our other previous work found that the goal of game bot parties is different from that of human users parties, and proposed a party log-based detection method~\cite{bib9}. This approach is however limited to detecting misbehavior in party play and cannot detect misbehavior in single play games. 

Gold farming group analysis uses the virtual economy in online games and traces abnormal trade networks formed by gold farmers, merchants, bankers, and buyers. To characterize each player, Itsuki \etal \cite{bib10} used four types of statistics: total action count, activity time, total chat count, and the amount of virtual currency managed in a given period of time. 
Seo \etal \cite{bib11} analyzed gold farming group connection patterns using routing and source location information.
Kwon \etal \cite{bib12} investigated gold farming networks and detected the entire network structure of gold farming groups. This work, while distantly related, is not concerned with the detection of bots, but with understanding the unique roles each bot plays in the virtual underground ecosystem given a valid detection. 

Sequence analysis uses iterated sequence datasets from login to logout.
Ahmed \etal \cite{bib13} studied activity sequence features, defined as the number of times a given player engages in an activity, such as the number of monsters killed and the number of times the player was killed. 
Kwon \etal \cite{bib14} used the combat sequence each avatar produces. 
Lee \etal \cite{bib15} examined the full action sequence of users on big data analysis platform. While such technique has been shown to work in the past, such feature lacks context, and might be easily manipulated by bot settings. 

Similarity analysis uses the fact that game bots have a strong regular pattern because they play to earn in-game money. Kwon \etal \cite{bib16} derived vectors using the frequency of each event and calculated the vector's cosine similarity with a unit vector. Game bots repeatedly do the same series of actions, therefore their action sequences have high self-similarity. Lee \etal \cite{bib24} employed self-similarity measures to detect game bots. They proposed the self-similarity measure and tested it in three major MMORPGs (``Lineage'', ``Aion'' and ``Blade\&Soul''). Their scheme requires a lot of data of certain behavior for establishing self-similarity.

Moving path analysis uses the fact that game bots have pre-scheduled moving paths, whereas human users have various moving patterns. 
Thawonmas \etal \cite{bib17} provided a method for detecting landmarks from user traces using the weighted entropy of the distribution of visiting users in a game map. They presented user clusters based on transition probabilities. 
To identify game bots and human users, Van Kesteren \etal \cite{bib18} took advantage of the difference in their movement patterns. 
Mitterhofer \etal \cite{bib19} detected the players controlled by a script with repeated movement patterns. 
Pao \etal \cite{bib20} used the entropy values of a user's trace and a series of location coordinates. They employed a Markov chain model to describe the behavior of the target trajectory. 
Pao \etal \cite{bib21} applied their method to various types of trajectories, including handwriting, mouse, and game traces, in addition to the traces of animal movement. However, their feature also can be evaded and noised by adaptive bots that integrate human-like moving behavior.

\section{Methods}
\label{sec:methods}

Before elaborating on the framework and workflow of our method, we first highlight the dataset and ethnical guidelines used for obtaining and analyzing it. 


\BfPara{Dataset} To perform this study, we rely on a real-world dataset obtained from the operation of Aion, a popular game.  Our Aion dataset contains all in-game action logs for 88 days, between April 9th and July 5th of 2010. During this period, there were 49,739 characters that played more than three hours. Among these players, 7,702 characters were game bots, identified and labeled by the game company. The banned list was provided by the game company to serve as the ground truth, and each banned user has been vetted and verified by human labor and active monitoring.

\BfPara{Ethnical and privacy considerations} In order to perform this study we follow best practices in ensuring users privacy and complying with ethical guidelines. First, the privacy of users in the data is ensured by anonymizing all personal identifiable information. Furthermore, consent of users is taken into account by ensuring that data analysis is within the scope of end user license agreement (EULA): upon joining Aion, users grant NCSoft, Inc. the full permission to use and share user data for analysis purpose with parties of NCSoft's choosing. One of such parties was our research group, and for research purpose only.

\subsection{Framework and workflow}
\begin{figure}[htb]
\includegraphics[width=0.47\textwidth]{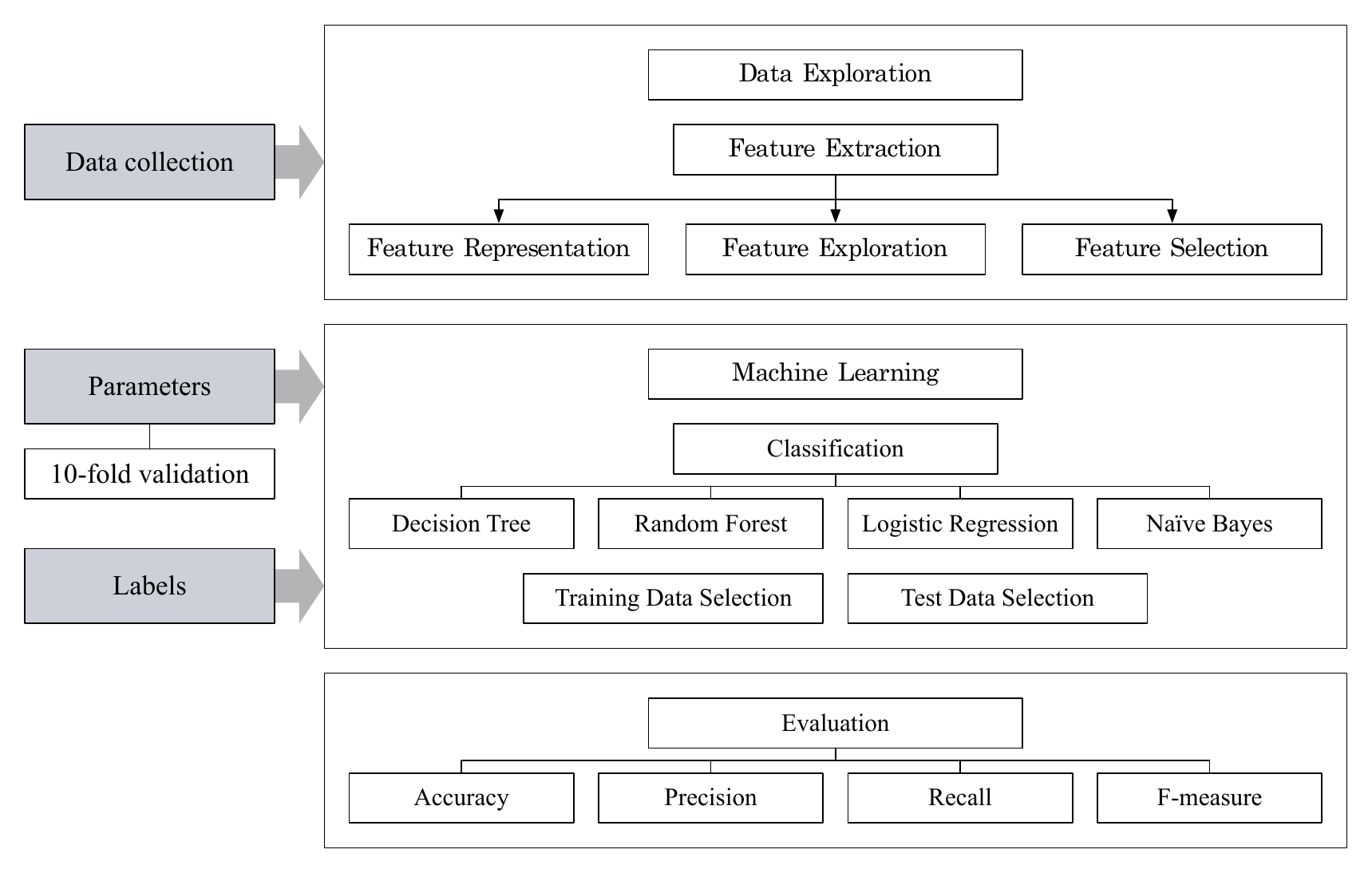}
\caption{Game bot detection framework based on user behavioral characteristics.}
\label{fig1}
\end{figure}

Our proposed framework for game bot detection is shown in Figure \ref{fig1}. We posed the problem of identifying game bots as a binary classification problem. At a high-level, our method starts with a data collection phase, followed by a data exploration phase (including feature extraction), a machine learning phase, and a validation phase. In the following we highlight each of those phases. 

\BfPara{Data collection} In the data collection phase, we gathered a dataset that combines in-game logs and chat contents. 

\BfPara{Data exploration} We then performed data exploration in order to comprehend the characteristics of the dataset using data preprocessing, feature extraction, feature representation, exploration, and selection for best discriminating between bots and normal users. In the feature representation procedure, we followed standard methods for unifying data and reducing its dimensionality. For example, we quantized each network measure into three clusters with low, medium, and high values using the k-means clustering algorithm.  In the feature exploration phase, we selected the components of the data vectors and pre-pocessed them. For example, we determined seven activities as social interactions and quantified the diversity of social interactions by the Shannon diversity entropy. In the feature selection phase, we selected significant features with the best-first search, greedy-stepwise search, and information gain ranking filter to avoid overfitting and reduce the features (thus improving the performance).

\BfPara{Machine learning} In the machine learning phase, we choose algorithms (e.g., decision tree, random forest, logistic regression, and na\"ive Bayes) and parameters (e.g., $k$-fold-cross validation parameters, specific algorithm parameters, etc.), and feed the data collected using the selected features in their corresponding representation. We further build models (using the data fed) and establish baselines by computing various performance metrics. 

 
\BfPara{Evaluation} In the evaluation phase, we summarize the performance of each classifier with the banned account list provided by the game company as a ground truth, by providing various performance measures, such as the accuracy, precision, recall, and F-measure.

\BfPara{Used features and their gap} As indicated in Table \ref{table2}, we classified the features we used in our work into personal and social features. Given that the aim of game bots is to earn unfair profits, there is a gap between the values of the personal features of game bots and those of human users. The personal features can be also categorized into player information and actions. The player information features include login frequency, play time, game money, and number of IP address. The player action features contain sitting (an action taken by players to recover their health), earning experience points, obtaining items, earning game money, earning player kill (PK) points, harvesting items, resurrecting, restoring experience points, being killed by a non-player and/or player character (NPC/PC), and using portals. The frequency and ratio of these actions reflects the behavioral characteristics of game bots and human users. For example, game bots sit more frequently than human users to recover health and mana points. Moreover, a player can acquire PK points by defeating players of opposing factions. PK points can be used to purchase various items from vendors. PK points are also used to determine a player's rank within the game world. In Aion, the more PK points a player has, the higher is the player's rank. The high ranking player can feel a sense of accomplishment. On the other hand, it is seen that game bots are not interested in rank.

\begin{table*}[t]
\begin{center}
\caption{Personal and social features.}
{\scriptsize
\begin{tabular}{p{2cm}p{3.8cm}p{11.1cm}}\hline
\multicolumn{2}{l}{Category} & Key idea\\ \hline
Personal feature & Player information & Login frequency, play time, game money, number of IP address\\ \hline
 & Player actions & Sitting, earning experience points, obtaining items, earning game money, earning player kill points, harvesting items, resurrection, restoring experience points, being killed by a non-player and/or player character, using portals\\ \hline
Social feature & Group activities & Party play time, guild activities\\ \hline
 & Social interaction diversity & Party play, friendship, trade, whisper, mail, shop, guild\\ \hline
 & Network measures & Degree centrality, betweenness centrality, closeness centrality, eigenvector centrality, eccentricity, authority, hub, PageRank, clustering coefficient\\ \hline
\end{tabular}
\label{table2}
}
\end{center}
\end{table*}

In addition, there is gap between the values of the social features of game bots and those of human users because game bots do not attempt to social as humans. The social features can be categorized into group activities, social interaction diversity, and network measures. The features of group activities include the average duration of party play and number of guild activities. Party play is a group play formed by two or more players in order to undertake quests or missions together. The goals of party play commonly are to complete difficult quests by collaboration and enjoy socialization. Interestingly, some game bots perform party play, but the goal of party play of the game bots is different from that of human users. Their aim is to acquire game money and items faster and more efficiently. Hence, there are the behavioral differences between game bots and human users. The social interaction diversity feature indicates the entropy of party play, friendship, trade, whisper, mail, shop, and guild actions. Game bots concentrate only on particular actions, whereas human users execute multiple tasks as needed to thrive in the online game world. The player's social interaction network can be represented as a graph with characters as the nodes and interactions between them as the edges. An edge between two nodes (players) in this graph may, for example, highlight the transfer of an item between the two nodes. The features of network measures include the degree, betweenness, closeness, eigenvector centrality, eccentricity, authority, hub, PageRank, and clustering coefficient. The definitions of the network measures are listed in Table \ref{table3}.

\begin{table*}[h]
\begin{center}
\caption{Definition of network measures. Network measures include degree, betweenness, closeness centrality, and efficiency.}
{\scriptsize
\begin{tabular}{lp{13cm}}
\hline
Network measures & Definitions\\ \hline
Degree centrality & The most intuitive notion of centrality focuses on the degree. The more edges an actor has, the more important it is.\\ \hline
Betweenness centrality & Counts the number of shortest paths between two nodes on which a given actor resides.\\ \hline
Closeness centrality & An actor is considered important if it is relatively close to all other actors. Closeness is based on the inverse of the distance of each actor to every other actor in the network.\\ \hline
Eigenvector centrality & Indicates that a given node has a relationship with other valuable nodes. A high eigenvector value for an actor means that a node has several neighbors with high eigenvector values.\\ \hline
Eccentricity & The eccentricity of node v is calculated by computing the shortest path between node v and all other nodes in the graph; then the longest shortest path is chosen.\\ \hline
Authority & Exhibits a node pointed to by many good hubs.\\ \hline
Hub & Exhibits a node that points to many good authorities.\\ \hline
PageRank & Assigns a numerical weight to each element of a hyperlinked set of documents, such as the World Wide Web, with the purpose of ``measuring'' its relative importance within the set.\\ \hline
Clustering coefficient & Quantifies how close neighbors are to being a clique. A clique is a subset of all of the edges connecting pairs of vertices of an undirected graph. \\ \hline
\end{tabular}
\label{table3}
}
\end{center}
\end{table*}

\section{Results and discussion}
In this section we review more concretely the behavioral characteristics of bots and humans based on the various features utilized, and using the aforementioned dataset. We then propose our bot detection mechanism based on discriminative features and by elaborating on details of the high level workflow in the previous section, including the performance evaluation. 

\subsection{Behavioral characteristics}
\subsubsection{Player information}
\begin{figure}[t]
\includegraphics[width=0.4\textwidth]{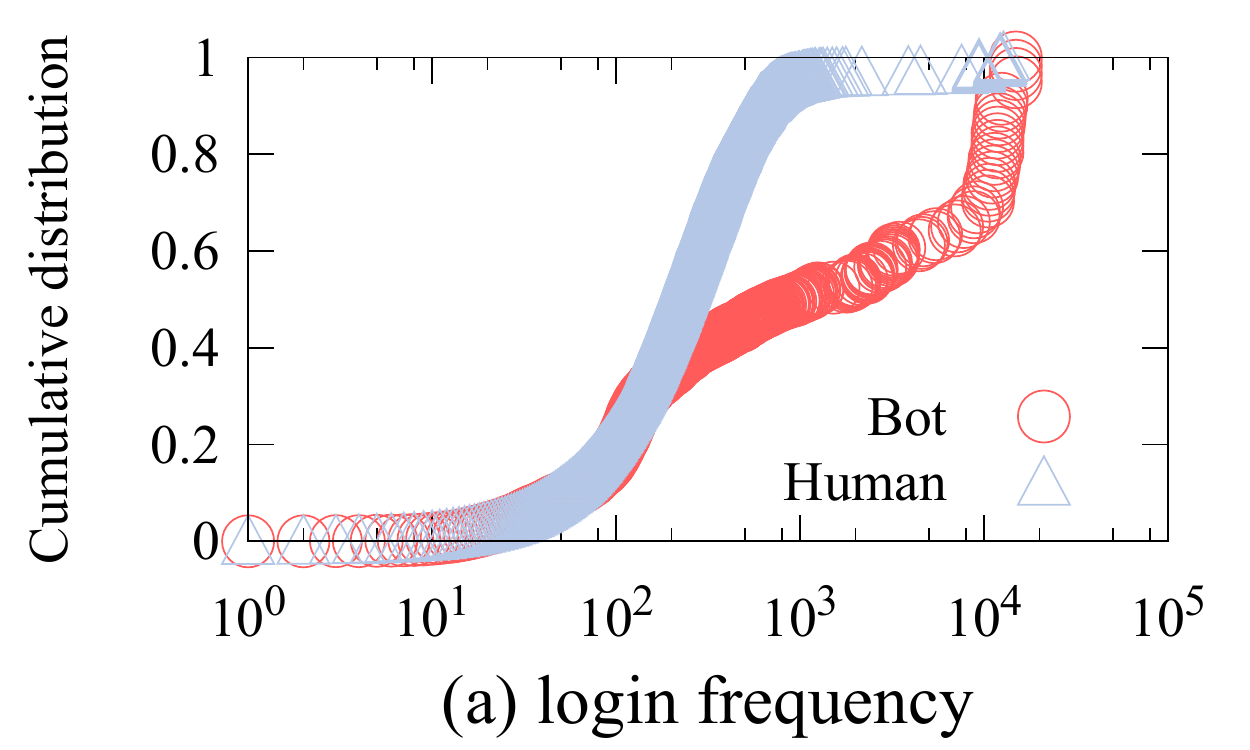}
\includegraphics[width=0.4\textwidth]{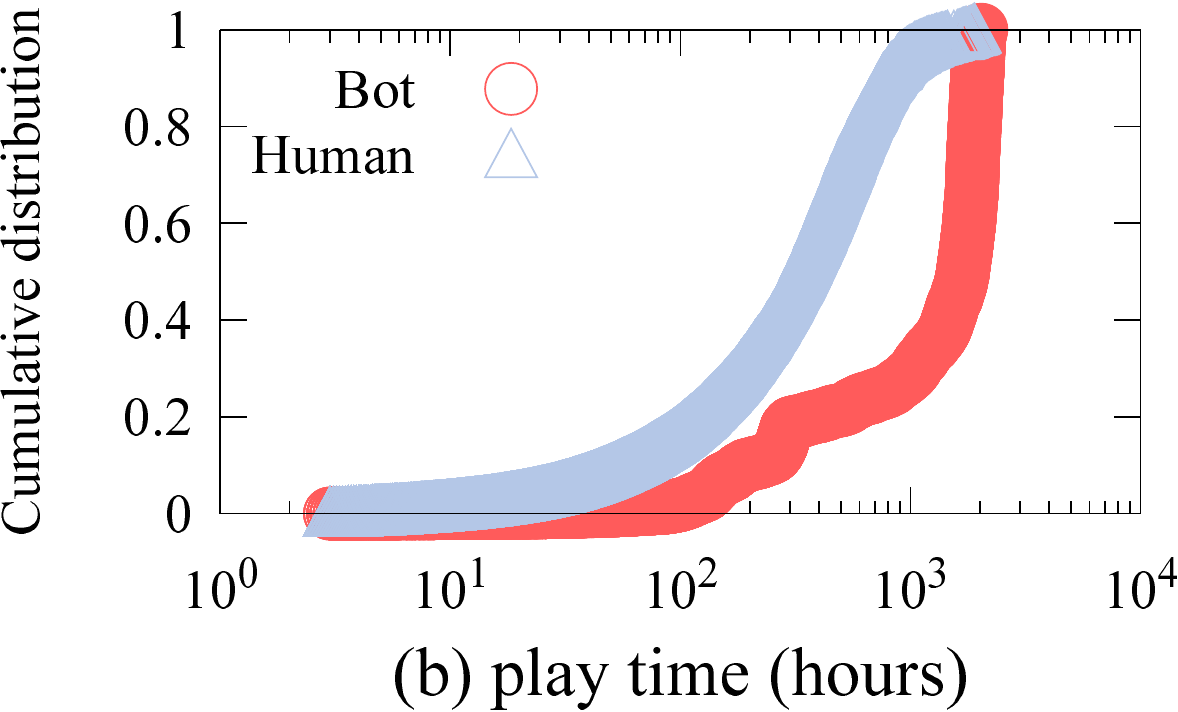}
\includegraphics[width=0.4\textwidth]{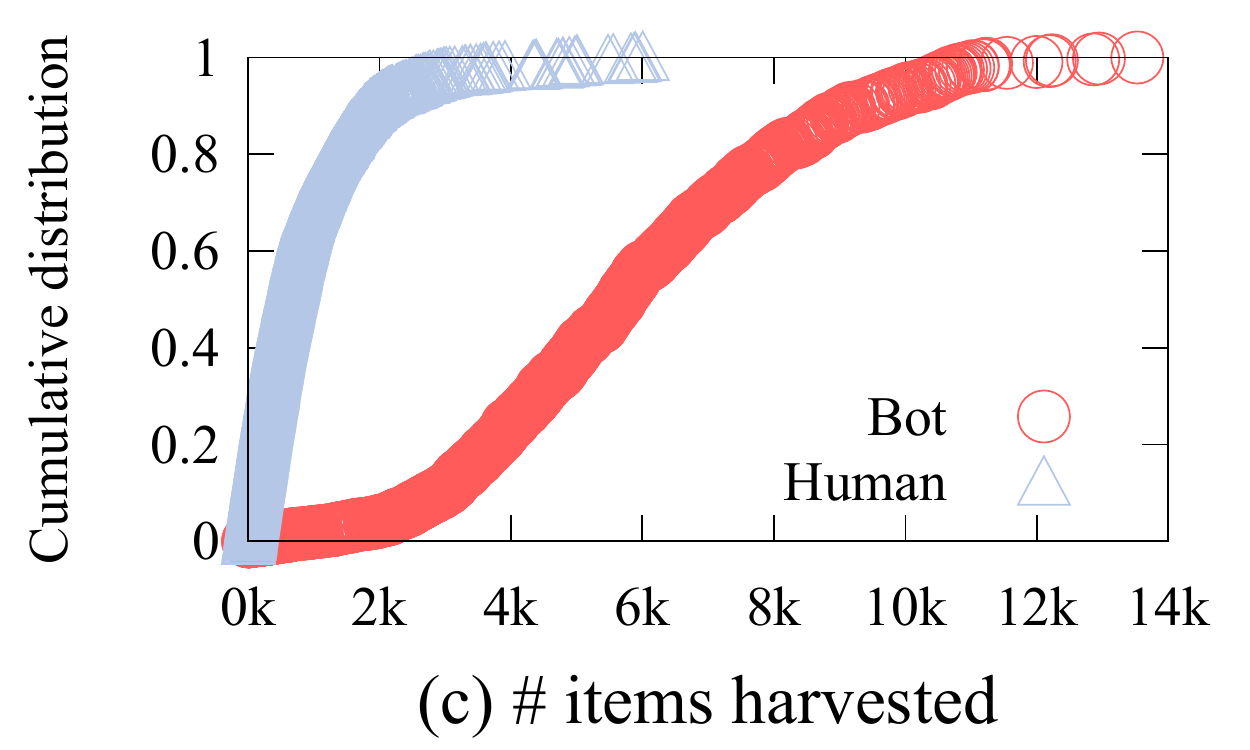}
\caption{{Player information.} (a) Cumulative distribution of the user login frequency. (b) Cumulative distribution of user play time. (c) Cumulative distribution of the number of items harvested by users.}
\label{fig2}
\end{figure}

We compared the distribution of player information features in order to identify the difference between the behavioral characteristics of game bots and human users more concretely. Figure \ref{fig2} shows how intensively game bots play games. Game bots often connect to the game and spend much longer time playing it than human users. Game bots can play a given game for 24 consecutive hours, whereas human users hardly connect to the game during working hours. Game bots invest significant time in a game until they are blocked. Figure \ref{fig2}(c) shows the cumulative distribution of the maximum number of items harvested by users per day. It is almost impossible for human users to harvest more than 1,000 items per day. Since this is repetitive and hard work, human users are easily exhausted. Nevertheless, 60\% of game bots harvest more than 5,000 items a day. This is an obvious characteristic for identifying game bots that we include in our feature set.

\begin{figure}[htb]
\includegraphics[width=0.45\textwidth]{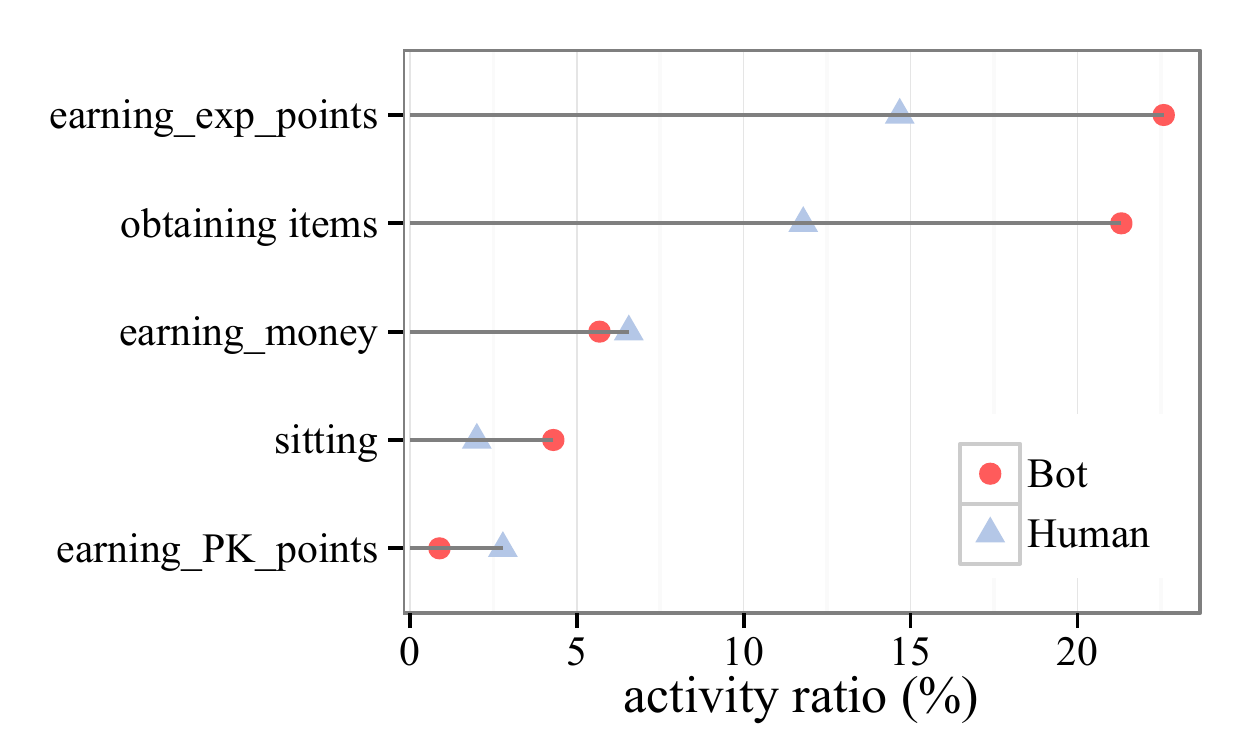}
\caption{{Comparison of activity ratios between game bots and human users.} The ratios of ``earning experience points'' and ``obtaining items'' of game bots are much higher than those of human users.}
\label{fig3}
\end{figure}

\subsubsection{Player actions}
We examined the frequency and ratio of player actions to determine the unique characteristics of game bots. Figure \ref{fig3} presents the ratios of the activities of both game bots and human users. The points in red indicate game bots, and those in blue indicate human users. The ratio of ``earning game money'' of game bots is nearly similar to that of human users. Remarkably, the ratios of ``earning experience points'' and ``obtaining items'' of game bots are much higher than those of human users. The cumulative ratio of ``earning experience points'', ``obtaining items'', and ``earning game money'' of game bots is close to 0.5, whereas that of human users is only 0.33. This implies that game bots concentrate heavily on profit-related activities, and human users enjoy various activities. In contrast, the ratio of ``earning PK points'' of human users is as much as three times that of game bots. This reflects the fact that game bots are not interested in rankings.

\subsubsection{Group activities}
\begin{figure}[htb]
\includegraphics[width=0.5\textwidth]{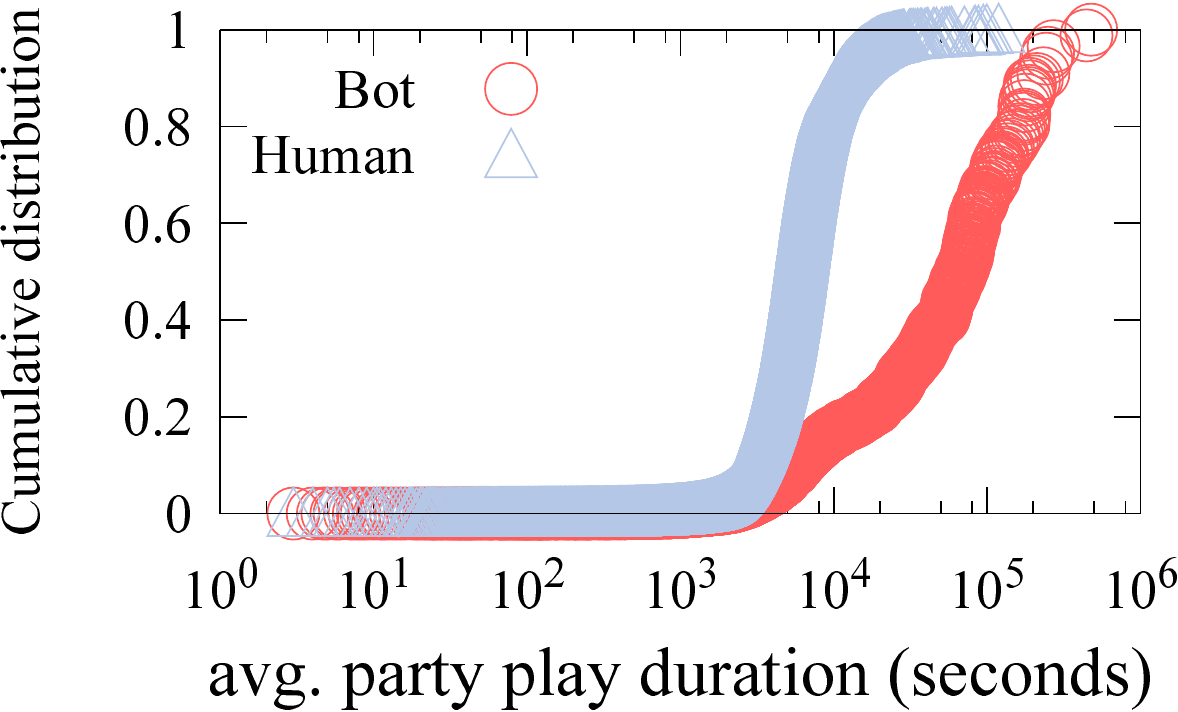}
\caption{{Cumulative distribution of user average party play time.} A total of 80\% of game bots last longer than 4 hours 10 minutes, whereas 80\% of human users last less than 2 hours 20 minutes.}
\label{fig4}
\end{figure}

Figure \ref{fig4} shows the distribution of the average party play time of game bots and human users. To acquire game money and items, some game bots form a party with other game bots. They can help each other not to be killed by monsters during party play. Consequently, their party play patterns are unusual. A total of 80\% of game bots last longer than 4 hours 10 minutes, whereas 80\% of human users last less than 2 hours 20 minutes. Since difficult missions can normally be completed within two hours through collaboration, human users do not maintain party play as long as game bots. 

\subsubsection{Social interaction diversity}
\begin{figure}[htb]
\includegraphics[width=0.5\textwidth]{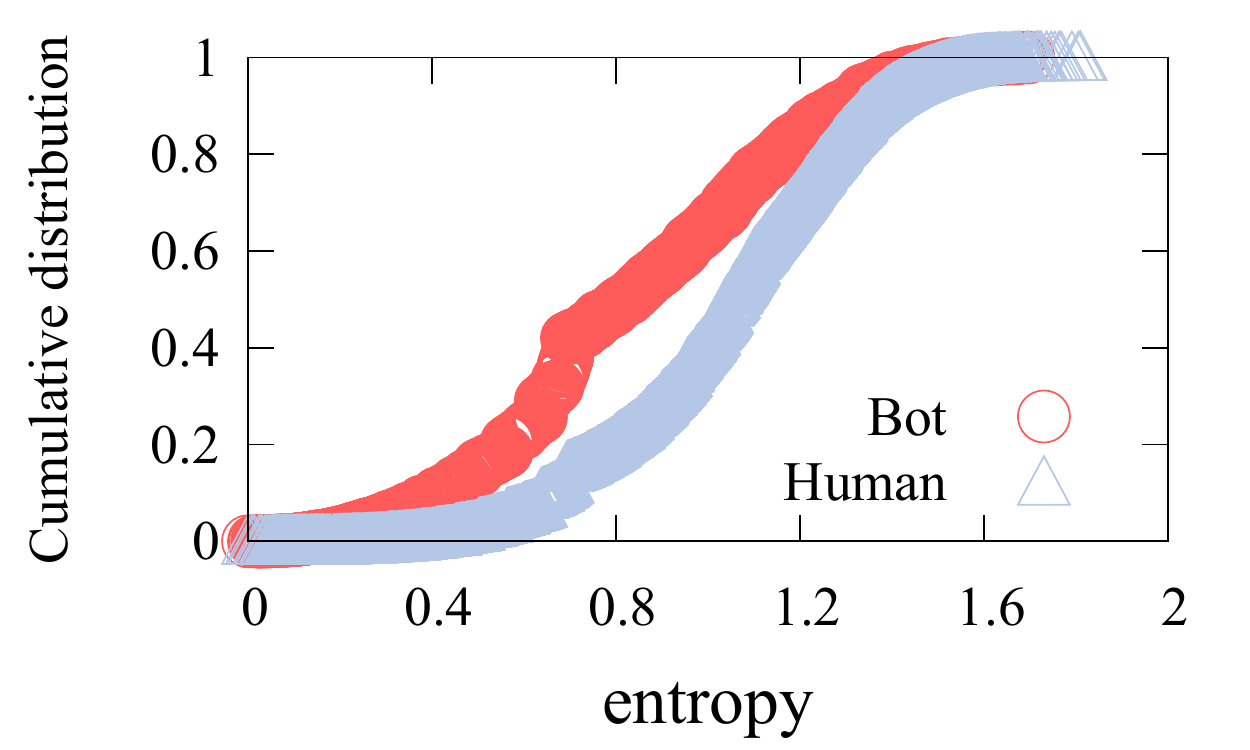}
\caption{{Cumulative distribution of user social interaction diversity.} The average entropy of game bot social interaction is much lower than that of human users (0.4299 and 0.8352, respectively).}
\label{fig5}
\end{figure}

Figure \ref{fig5} shows the cumulative distribution of the entropy of social interactions. First, we determined seven activities as social interactions: party, friendship, trade, whisper, mail, shop, and guild. We quantified the diversity of social interactions by calculating the Shannon diversity entropy defined by:
\begin{eqnarray}\label{eq:shannon_diversity} 
H' = -\sum_{i=1}^{n}{p_i \ln{p_i}}
\end{eqnarray}
\begin{conditions}
n & number of social interaction types\\
p_i & relative proportion of the $i^{th}$ social interaction type\\
\end{conditions}

The entropy of the social interactions of a player indicates the various activities performed by the player. Figure \ref{fig5} represents the fact that human users enjoy diverse activities, whereas game bots do not. We notice that game bots are interested in other activities.

\subsubsection{Network measures}
In Figure \ref{table4}, we present the basic directed characteristics of each network of the game bot and human groups from Aion~\cite{bib22}. First, we see that the average degree of the human group is approximately 18 times larger compared with the game bot group in the party network. The reason is that human users form a party with many and unspecified users, whereas game bots play with several specific other game bots. The average degree of the friendship network of the human group is larger by a factor of approximately four compared with the game bot group. This fact indicates that the friendship of game bots is utterly different from that of human users. Game bot friends simply mean other game bots with which to play. The fact that the average degree of the human group is 2.5 times larger than the game bot group is observed in the case of the trade network. However, the average clustering coefficient of the game bot group is approximately five times larger compared with the human group. We assume that game bots have roles~\cite{bib12, bib13}. For instance, some game bots are responsible for gold farming, while other game bots gather game money and items from gold farmers or sell them for real money~\cite{woo2011can}. 

Interestingly, in the case of the mail network of the game bots, we discovered nine spammers during the observation period. The number of mail pieces sent by the spammers is 1,000 times per person on average. We observed the behavioral characteristics of the spammers in more detail. Hence, we found that they only send mail and stay online for a short period of time in the  online game world. 

We also observed the existence of five collectors who received items attached to mail from many other game bots. These collectors received items over 6,000 times during the observation period. This shows that there are several gold farming groups. In the case of the shop network, we can see the smallest number of nodes of both groups. Players are immobile in the merchant mode, and thus cannot engage in any action that requires movement, such as hunting monsters, harvesting items, etc. Consequently, game bots do not focus on the merchant mode because it can be a waste of time for them.

\begin{table*}[t]
\caption{Basic network characteristics of six interaction networks. The average degree of all interaction networks of the human group is higher than that of the game bot group. This shows that game bots do not enjoy socializing with other users.}
\begin{tabular}{lrrrrrrrrrrrr} \hline
 & \multicolumn{2}{l}{{Party}} & \multicolumn{2}{l}{{Friendship}} & \multicolumn{2}{l}{{Trade}} & \multicolumn{2}{l}{{Whisper}} & \multicolumn{2}{l}{{Mail}} & \multicolumn{2}{l}{{Shop}} \\ \cline{2-13}
& \multicolumn{1}{c}{Bot} & \multicolumn{1}{c}{Human} & \multicolumn{1}{c}{Bot} & \multicolumn{1}{c}{Human} & \multicolumn{1}{c}{Bot} & \multicolumn{1}{c}{Human} & \multicolumn{1}{c}{Bot} & \multicolumn{1}{c}{Human} & \multicolumn{1}{c}{Bot} & \multicolumn{1}{c}{Human} & \multicolumn{1}{c}{Bot} & \multicolumn{1}{c}{Human} \\ \hline
Nodes & 1756 & 33924 & 479 & 24628 & 4003 & 30640 & 434 & 16209 & 4848 & 28362 & 305 & 7001\\ 
Edges & 2463 & 862021 & 749 & 174626 & 9809 & 162236 & 656 & 248133 & 12873 & 76844 & 362 & 11824\\ 
Avg. degree & 1.4 & 25.41 & 1.56 & 7.09 & 2.45 & 5.29 & 1.51 & 15.31 & 2.66 & 2.71 & 1.19 & 1.7\\ 
Network diam. & 22 & 15 & 9 & 15 & 25 & 18 & 23 & 12 & 9 & 24 & 5 & 28\\ 
Avg. C.C. & 0.1 & 0.07 & 0.07 & 0.09 & 0.41 & 0.08 & 0.01 & 0.05 & 0.12 & 0.19 & 0.12 & 0.01\\ 
Avg. path len. & 6.14 & 3.77 & 2.18 & 4.7 & 5.66 & 5.41 & 6.41 & 3.65 & 2.16 & 7.55 & 1.58 & 8.14\\ \hline
\end{tabular}
\label{table4}
\end{table*}

\subsubsection{The triad census}
\begin{figure}[t!]
\centering
\subfigure{\label{fig:fig6_1}\includegraphics[width=0.40 \textwidth]{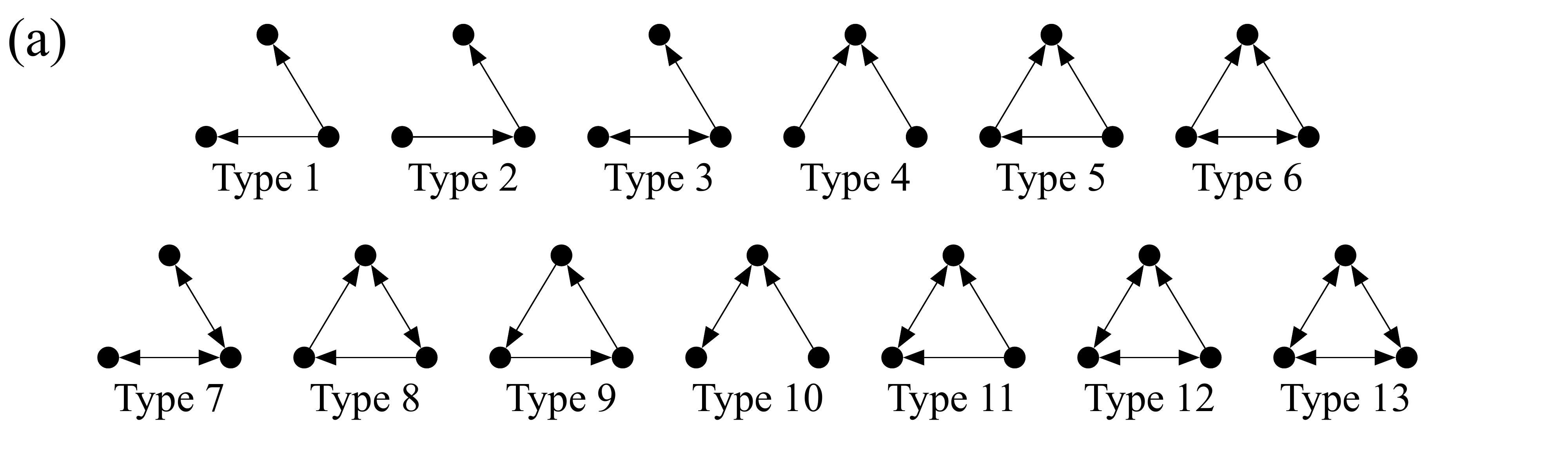}} 
\subfigure{\label{fig:fig6_2}\includegraphics[width=0.40 \textwidth]{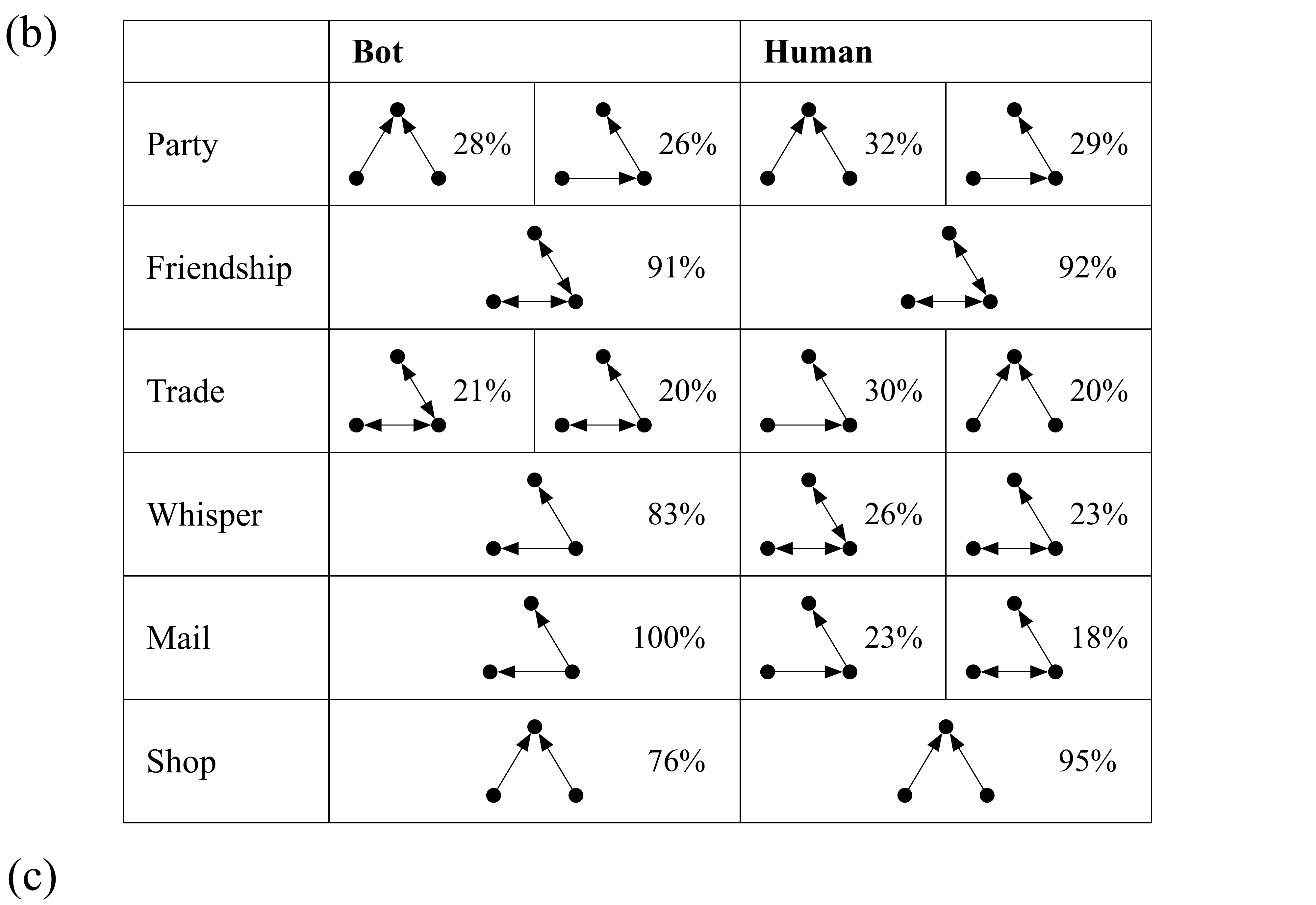}} 
\subfigure{\label{fig:fig6_2}\includegraphics[width=0.40 \textwidth]{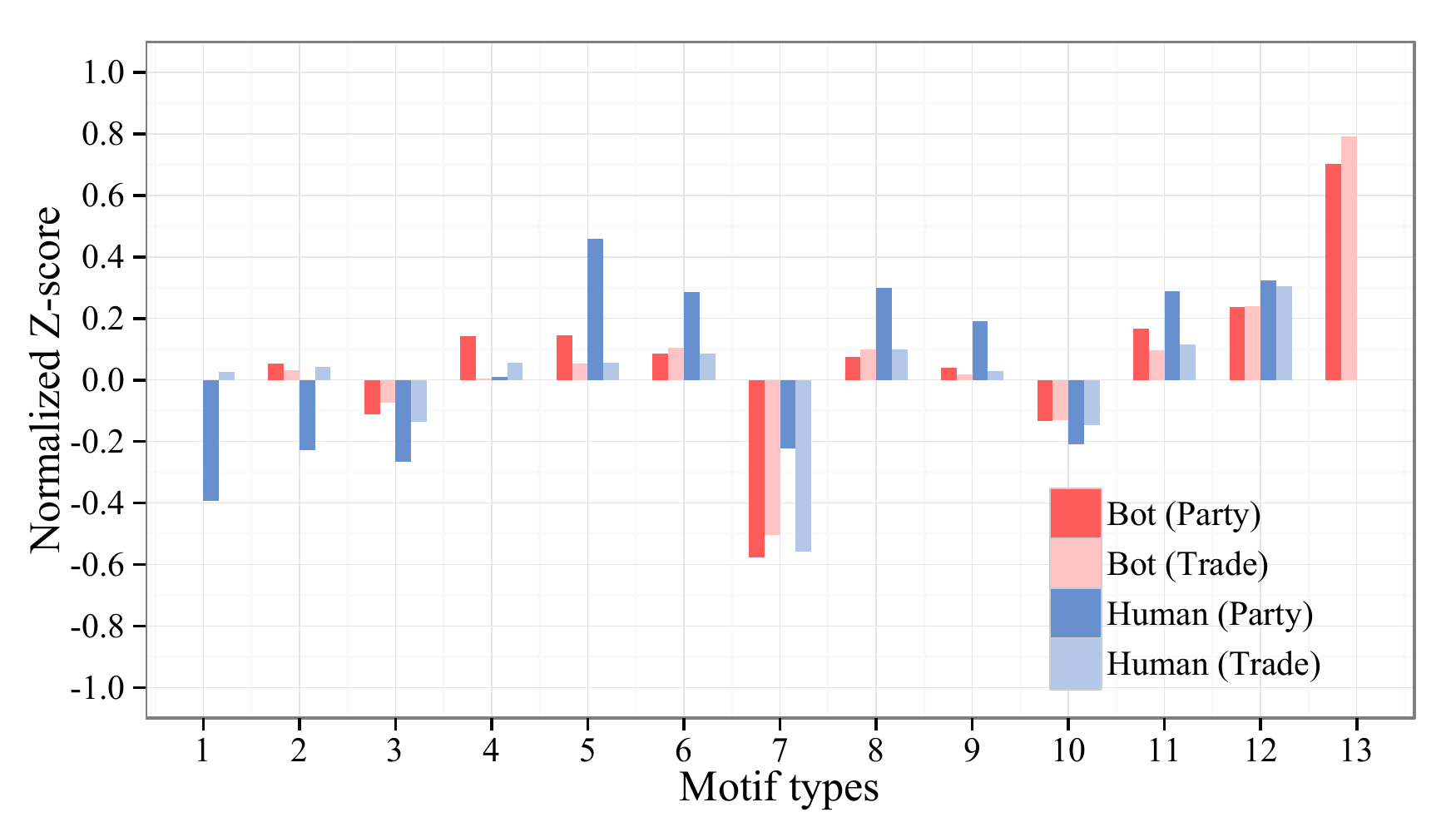}} 
\caption{{Network motif analysis of node triplets reveals detailed interaction patterns in directed networks of game bots and human users.} (a) The 13 possible motifs composed of three nodes in a directed network. (b) The fractions of each motif type in each of the six networks. Those motifs that account for fewer than 18\% of all motifs are not shown. Friendship, whisper, mail, and shop of the game bot group, and friendship and shop of the human group show one dominant motif each, consistent with the high or low reciprocity found in the networks. (c) A closer look at the (normalized) Z-score triad census of party and trade networks where no dominant motif is evident; the Z-score method is employed to determine significantly over and underrepresented triangular motifs.}
\label{fig6}
\end{figure}

The relative prevalence of each of the 13 triad network motifs given in Figure \ref{fig6}(a) indicates the interaction pattern in the networks in more detail~\cite{bib23}. For our Aion networks, we show the interaction pattern in Figure \ref{fig6}(b) in terms of both the fractions of each motif type and the Z-scores assessed against the null model (Eq.~\eqref{eq:zscore}, also see \ref{A2_Table} and \ref{A3_Table}). This score is defined as follows:

\begin{eqnarray}\label{eq:zscore} 
Z_i = {N_{i}^{\textnormal{real}} - N_{i}^{\textnormal{random}} \over \sigma_{i}, ^{\textnormal{random}}}
\end{eqnarray}
where $N_{i}^{\textnormal{real}}$ is the number of motif $i$ found observed in the network, 
$N_{i}^{\textnormal{random}}$ is the expected number in the randomized network, and 
$\sigma_{i}^{\textnormal{random}}$ is the standard deviation of its expected number in the randomized network.

\BfPara{Findings} Interestingly, the friendship, whisper, mail, and shop networks of the game bot group, and the friendship and shop networks of the human group, show one predominant motif type. For instance, in the friendship network, type 7 accounts for more than 90\% of the node triplet relationships, which can be attributed to the highly reciprocal nature of the interactions. The opposite reasoning can be applied to shop: low reciprocity reflects the existence of big merchants. Moreover, in the whisper and mail network of the game bot group, type 1 accounts for more than 80\% of the node triplet relationships. This reflects the fact that some game bots send information about the location coordinates of monsters to other game bots in the case of the whisper network.

Some game bots send several mail pieces in the case of the mail network. Comparing the prevalence of motifs against the null models allows us to detect signals discounted by random expectation, and this is done via the Z-scores (Eq.\eqref{eq:zscore}). This is particularly necessary and illuminating in the case of the other two networks (party and trade) because, by considering the null models, we can see that although multiple motifs can be similarly abundant (Figure \ref{fig6}(b)), some can be significantly over or underrepresented, as we can see in Figure \ref{fig6}. In the case of the human group, the overrepresented motif type 5 (with $\tilde{Z}$\textgreater0.4, the normalized version $\tilde{Z} \equiv Z_{i} \/ \sqrt{\displaystyle{\Sigma}_{i}({Z_{i}^{2}})}$) is indeed closed triangles, consistent with the relatively high clustering tendencies in the party network. In the case of the game bot group, the overrepresented motif type 13 shows the fact that there is a large gap between the number of motifs observed in the network and the expected number of motifs in the randomized network. This reflects the fact that game bots have their own group for helping and trading with each other.

\subsubsection{Network overlap}
\begin{figure}[t!]
\centering
\includegraphics[width=0.47\textwidth]{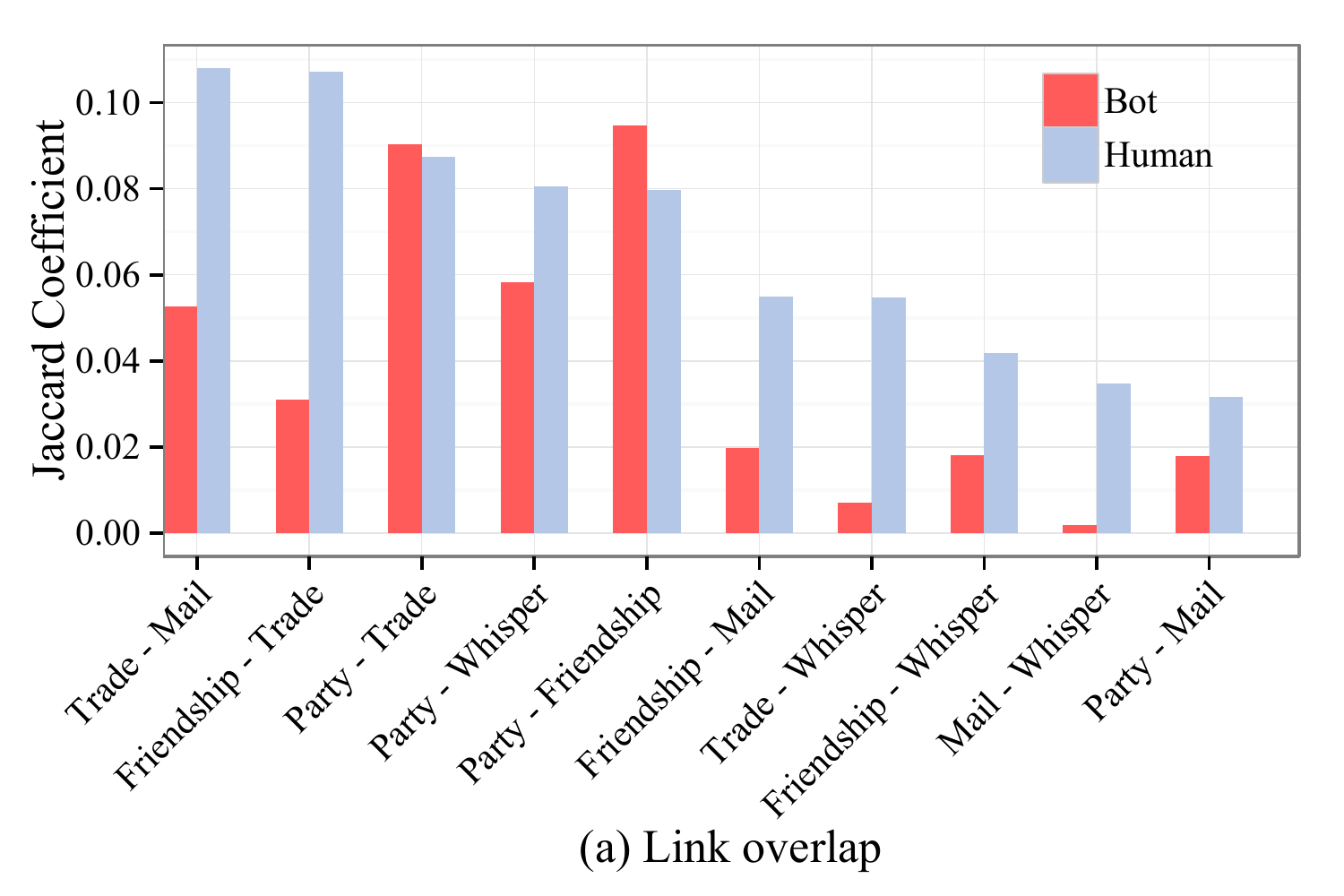}
\includegraphics[width=0.47\textwidth]{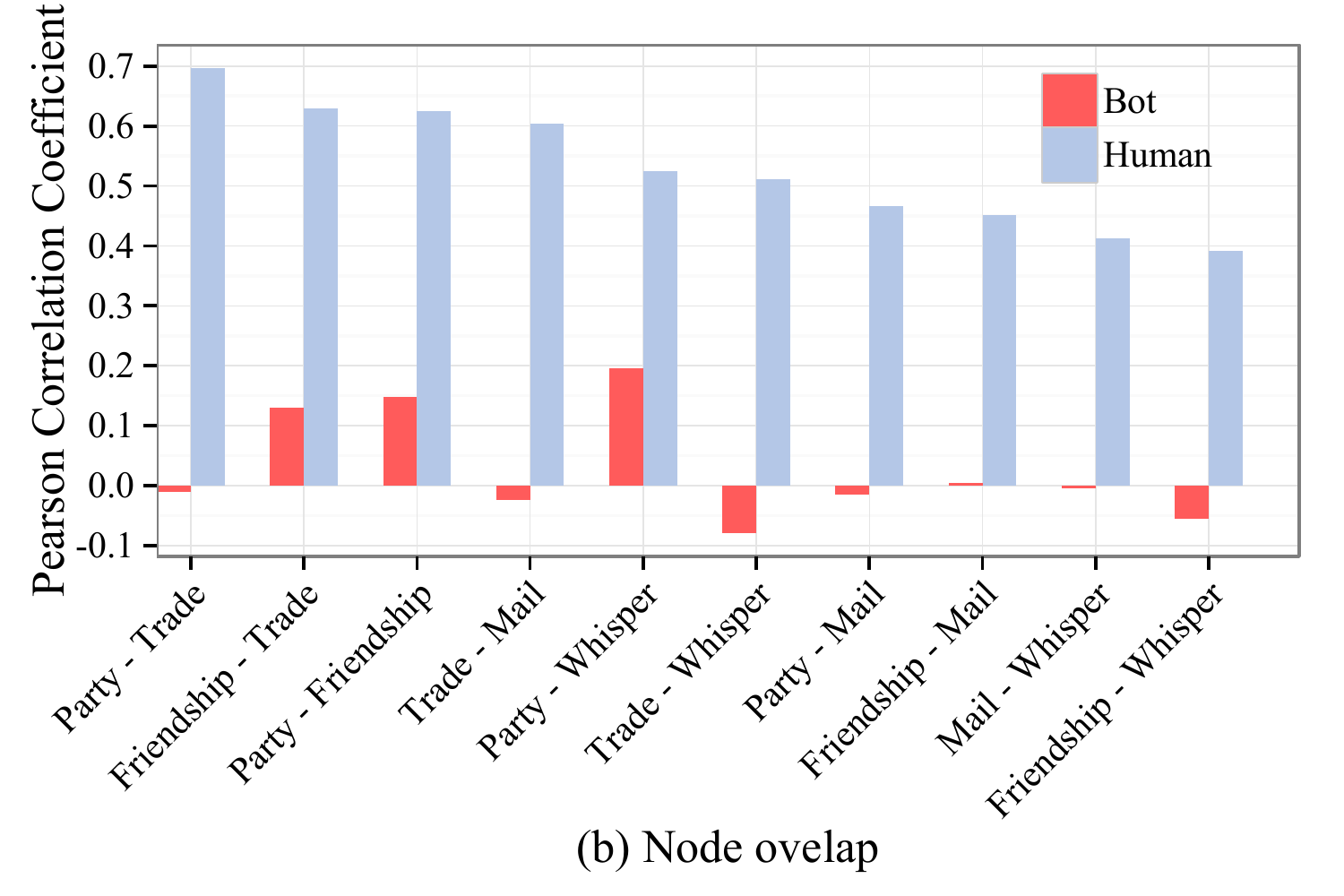}
\caption{{Pairwise network overlap indicates similarity or dependence between interactions.} (a) Link overlap. The game bot group has higher Jaccard coefficient in the party-friendship and party-trade pairwise networks. (b) Node overlap that quantifies the node degree overlap between different networks. The human group has high degree overlap between 0.4 and 0.7, whereas the game bot group has degree overlap lower than 0.2 in all networks.}
\label{fig7}
\end{figure}

To determine how pairwise networks are correlated, we studied the network similarities between the game bot and human groups. For example, two networks can show similar clustering values, and yet this does not guarantee at all that nodes connected in one network are connected in another, or that the nodes show similar levels of activity. Thus, we consider here two measures of network overlap. The first is the link overlap between two networks quantified by the Jaccard coefficient. The second is the degree overlap given by the Pearson correlation coefficient between node degrees in network pairs. The results of link and degree overlap for ten network pairs of the game bot and human groups are given in Figure \ref{fig7}. By examining the link overlap (Figure \ref{fig7} (a)), we found that the game bot group has higher Jaccard coefficient in the party-friendship and party-trade pairwise networks. This is a result of the fact that the main activities of game bots are party play and trading items. The friend list offers convenience to a game bot when it wants to form a party group. Game bots gather game money and items collected through party play in an account by trading. Then the account that collects the cyber assets changes the game money and items to real money.

Node degree overlap (Figure \ref{fig7} (b)) is another way of seeing the connection between interactions: here, for instance, the party-trade pairwise networks of the human group show a positive Pearson correlation coefficient value that exceeds 0.7, which can be understood by the fact that a party activity, being above all the favorite way of engaging in battles or hunting, often concludes with members trading booties. In contrast, the Pearson correlation coefficient values of the game bot group are extremely low because game bots maintain relationships with a small number of other game bots.

\subsection{Game bot detection}
We took a discriminative approach to learning the distinction between game bots and human users in order to detect the game bot and build automatic classifiers that can automatically recognize the distinction. We divided the dataset into training and test sets, built the classifiers through the training dataset, and evaluated the trained classifiers through the test dataset. In addition, we performed 10-fold cross-validation to avoid classifiers from being overfitted to the test data. Cross-validation generalizes the classifier trained by the test data to the validation data. 10-fold cross-validation divides the dataset into ten groups, trains the learning model with randomly selected nine groups, and verifies the classifiers from the model with one group. These training and validation processes are repeated ten times.

\subsubsection{Feature selection}
\begin{figure}[t]
\centering
\includegraphics[width=0.46\textwidth]{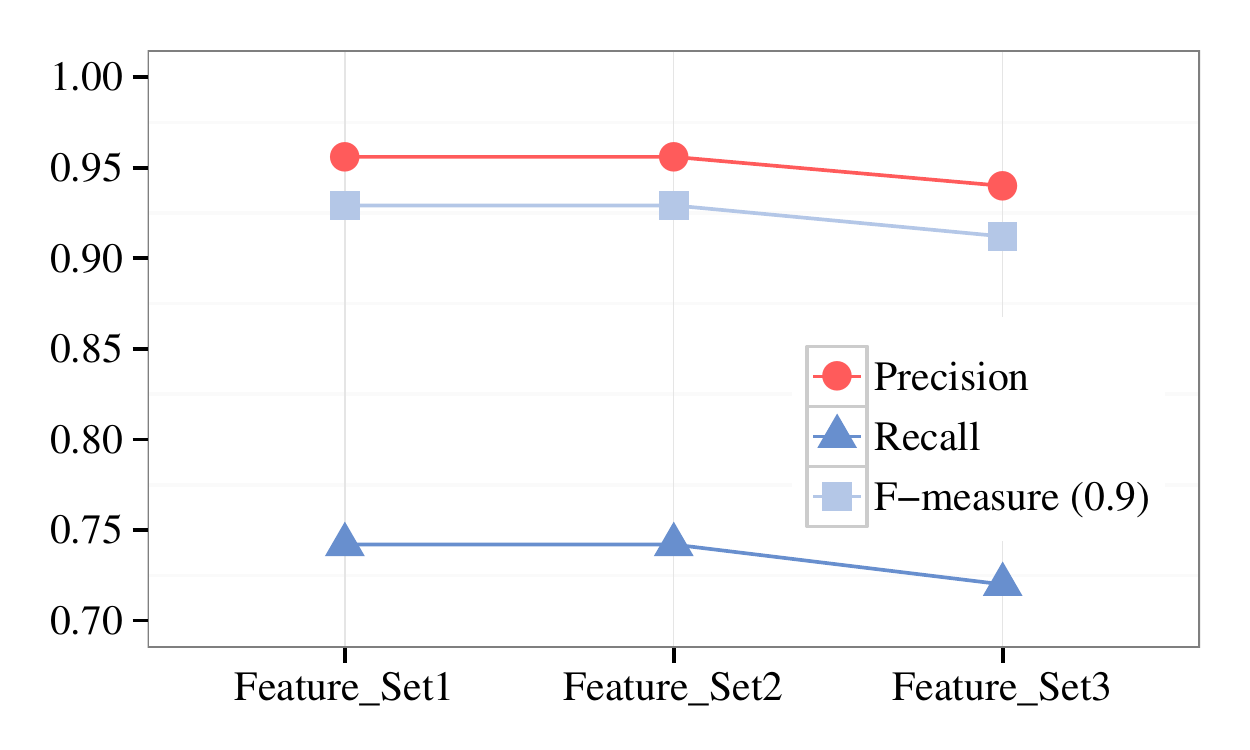}
\caption{{Performance comparison of feature sets.} Feature\_Set2 has as high performance as Feature\_Set1.}
\label{fig8}
\end{figure}

We compared the bot detection results from our model with the banned account list provided by the game company in order to evaluate the proposed framework upon running our detection method of selected features. We conducted feature selection with the best first, greedy stepwise, and information gain ranking filter algorithms in advance in order to improve the selection process. Feature\_Set1 consists of all the features (114) mentioned in section \ref{sec:methods}. Feature\_Set2 is composed of the top 62 features extracted by the information gain ranking filter algorithm. Feature\_Set3 is comprised of the six features selected by the best first and greedy stepwise algorithms. Figure \ref{fig8} shows the classification results using these three feature sets. Feature\_Set3 presents lower performance than Feature\_Set1 and Feature\_Set2. In comparison, Feature\_Set2 has almost the same performance as Feature\_Set1, although the number of Feature\_Set2 is barely half that of Feature\_Set1. Thus, we finally selected Feature\_Set2 for game bot detection.

\subsubsection{Classification and evaluation}
The results of the users' behavioral pattern analysis for game bot detection are listed in Table \ref{table5}. The four classifiers used as training algorithms---decision tree, random forest, logistic regression, and na\"ive Bayes---are tested on Feature\_Set2. The performances are listed in terms of overall accuracy, precision, recall, and F-measure. Random forest outperforms the other models. Its overall accuracy, precision value, recall value, and F-measure with emphasis on precision ($\alpha$ = 0.9) are 0.961, 0.956, 0.742, and 0.929, respectively. As can be seen, the recall value is slightly low. We analyzed the characteristics of true positive, false positive, false negative, and true negative cases to inquire into the cause of this phenomenon.

The random forest technique is a well-known ensemble learning method for classification and it constructs multiple decision trees in its training phase to overcome the decision tree's overfitting problem.  The random forest learning is also robust when training with imbalanced data set. It is also useful when training large data with a lot of features. Our data set consists of 85\% of human players and 15\% of game bots---so it is considered as an imbalanced and large data set---and random forests perform well in that context given that the context meets the settings in which random forests are to perform ideally. 

Na\"ive Bayes showed the lowest performance among four classifiers, and that is probably because of its nature as a generative model that requires independence of features.  Although we performed feature selection, still there are correlations between selected features used in our experiment. For example, obtaining\_items\_count, earning\_exp\_points\_count, harvesting\_items\_max\_count, party\_eccentricity, play\_time and obtaining\_items\_ratio are less significant features. However, those features are also naturally correlated and they cannot be easily separated because they are all related to essential game behaviors (hunting, harvesting, collaboration, etc., which are all related to high level process). Indeed, such hypothesis is confirmed by removing those features, bringing the performance of the na\"ive Bayes on par with other algorithms.

\begin{table*}[t]
\begin{center}
\caption{Precision, recall, and F-measure (0.9) ratios for each classifier. The random forest model employs the highest performance with overall accuracy rate of 0.961.}
{
\begin{tabular}{lrrrrrrr}
\hline
\multicolumn{1}{l}{\multirow{2}{*}{\textbf{Classifier}}} & \multicolumn{1}{l}{\multirow{2}{*}{\textbf{\begin{tabular}[c]{@{}l@{}}Overall\\ Accuracy\end{tabular}}}} & \multicolumn{3}{l}{\textbf{Human}} & \multicolumn{3}{l}{\textbf{Bot}} \\ \cline{3-8} 
\multicolumn{1}{l}{} & \multicolumn{1}{l}{} & \multicolumn{1}{c}{Precision} & \multicolumn{1}{c}{Recall} & \multicolumn{1}{c}{F-Meas.(0.9)} & \multicolumn{1}{c}{Precision} & \multicolumn{1}{c}{Recall} & \multicolumn{1}{c}{F-Meas.(0.9)} \\ \hline
Decision Tree & 0.955 & 0.96 & 0.989 & 0.963 & 0.911 & 0.737 & 0.89\\
Random Forest & 0.961 & 0.961 & 0.995 & 0.964 & 0.956 & 0.742 & 0.929\\
Logistic Regression & 0.955 & 0.956 & 0.994 & 0.96 & 0.95 & 0.705 & 0.918\\
Na\"ive Bayes & 0.948 & 0.96 & 0.981 & 0.962 & 0.859 & 0.734 & 0.845\\ \hline
\end{tabular}
\label{table5}
}
\end{center}
\end{table*}

\begin{figure}[t!]
\centering
\includegraphics[width=0.49\textwidth]{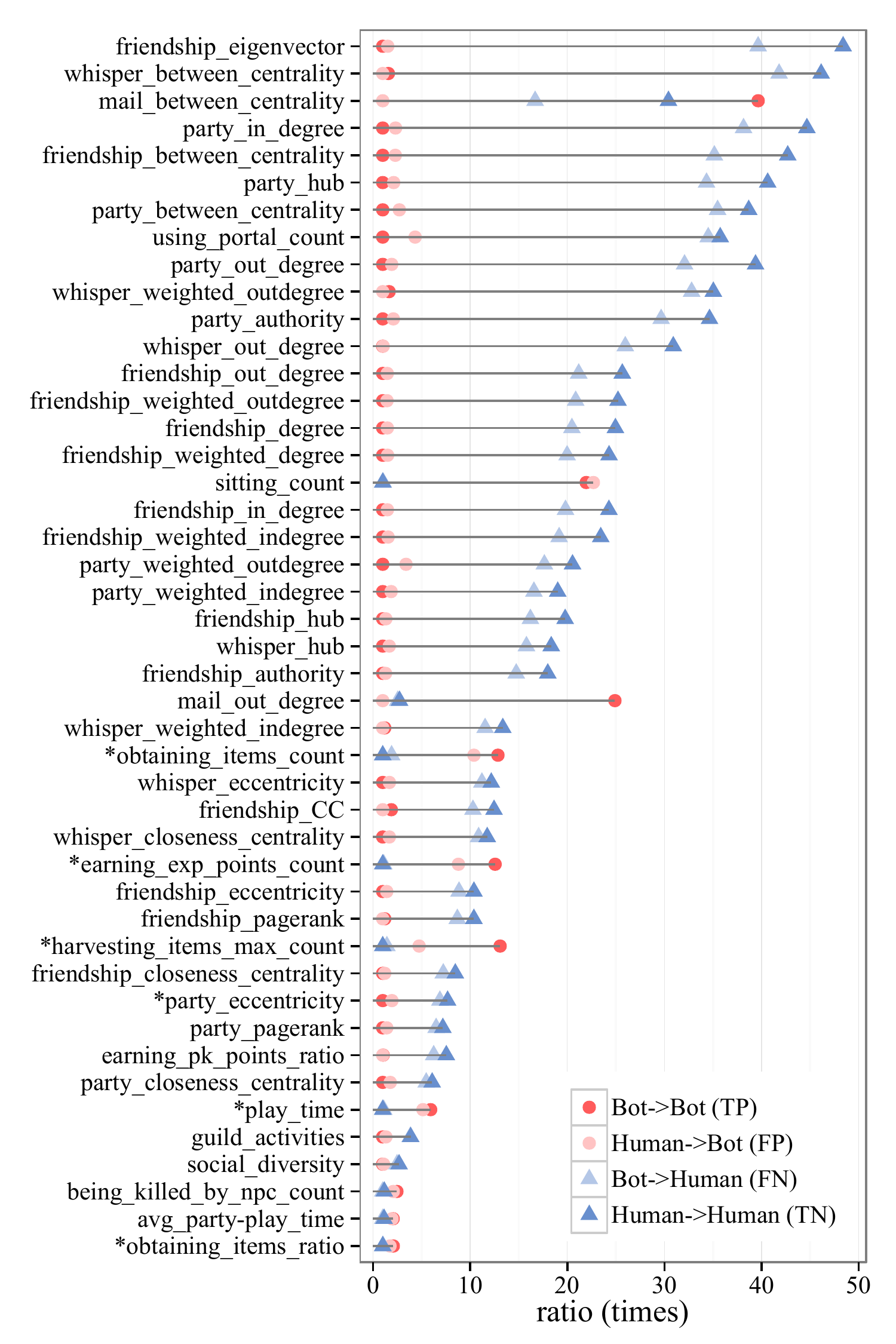}
\caption{{Comparison of four cases: true-positive, false-positive, false-negative, and true-negative.} The ratios of false-positive cases are exceedingly similar to those of true-positive cases. The ratios of false-negative cases are similar to those of true-negative cases. }
\label{fig9}
\end{figure}

Figure \ref{fig9} shows the relative similarities and differences of the classification evaluation outcomes (classes): true positive, false positive, false negative, and true negative. 
To obtain the relative similarity, we normalize all classes by the lowest class value, thus comparing outcomes relatively. Such normalization would bring the lowest class in the evaluation to one. For each class other than the lowest, we calculated the ratio by dividing the values of the other classes by the value of the lowest class. The pattern of the relative similarity is consistent for most features and classes, with the exception of the ``mail\_between\_centrality'' and ``mail\_outdegree'' features. It is highly probable that game bots had not been detected yet in the case of false negatives. This also implies that human users temporarily employed a game bot in the case of false positives. To confirm this observation, we analyzed the case of false positives weekly and finally found harvesting and party play game bots.

\section{Conclusions}
We proposed a multimodal framework for detecting game bots in order to reduce damage to online game service providers and legitimate users. We observed the behavioral characteristics of game bots and found several unique and discriminative characteristics. We found that game bots execute repetitive tasks associated with earning unfair profits, they do not enjoy socializing with other players, are connected among themselves and exchange cyber assets with each other. Interestingly, some game bots use the mail function to collect cyber assets. We utilized those observations to build discriminative features. We evaluated the performance of the proposed framework based on highly accurate ground truth -- resulting from the banning of bots by the game company. The results showed that the framework can achieve detection accuracy of 0.961. Nonetheless, we should consider that the banned list does not include every game bot. 

The game company imposes a penalty point on an account that performs abnormal activities, and eventually blocks the account when its cumulative penalty score is quite high. Some game bots can evade the penalty scoring system of the game companies. Hence, the actions of a player are more important than whether the player is banned or not, and we concede that a player is a game bot when the player's actions are abnormal. We focused on those user behavioral patterns that reflect user status to interpret the false positive cases, and hypothesize that they are game bots not yet blocked, and false negative cases are human users occasionally employing a game bot. Although different from those in the banned list, they behave in the same pattern. We believe that our detection model is more robust by relying on multiple classes of features, and its analyses promise further interesting directions in understanding game bot and their detection.

\section{Acknowledgements}
This research was supported by Basic Science Research Program through the National Research Foundation of Korea (NRF) funded by the Ministry of Science, ICT \& Future Planning (2014R1A1A1006228).


\section{Appendix}
Complete frequency distribution for triangular motifs is shown in Table~\ref{A2_Table}.
Network diameters from 100 randomized network versions is shown in Table~\ref{A1_Table}.    The network diameters from 100 randomized network versions and a comparison between the bots and human users is shown in Table~\ref{A3_Table}.  

\begin{table*}[h]
\caption{Multimodal characteristics of the online game.}\label{A2_Table}
\begin{tabular}{lp{1.5cm}p{1.5cm}p{1.5cm}p{1.5cm}p{1.5cm}p{1.5cm}p{1.5cm}p{1.5cm}p{1.5cm}p{1.5cm}p{1.5cm}p{1.5cm}}
\hline
\multirow{2}{*}{} & \multicolumn{2}{l}{\textbf{Party}} & \multicolumn{2}{l}{\textbf{Friendship}} & \multicolumn{2}{l}{\textbf{Trade}} & \multicolumn{2}{l}{\textbf{Whisper}} & \multicolumn{2}{l}{\textbf{Mail}} & \multicolumn{2}{l}{\textbf{Shop}} \\ \cline{2-13} 
 & \multicolumn{1}{l}{Bot} & \multicolumn{1}{l}{Human} & \multicolumn{1}{l}{Bot} & \multicolumn{1}{l}{Human} & \multicolumn{1}{l}{Bot} & \multicolumn{1}{l}{Human} & \multicolumn{1}{l}{Bot} & \multicolumn{1}{l}{Human} & \multicolumn{1}{l}{Bot} & \multicolumn{1}{l}{Human} & \multicolumn{1}{l}{Bot} & \multicolumn{1}{l}{Human} \\ \hline

\textbf{Type 1} & \multicolumn{1}{r}{ 15.04 } & \multicolumn{1}{r}{ 17.78 } & \multicolumn{1}{r}{ 0.16 } & \multicolumn{1}{r}{ 0.56 } & \multicolumn{1}{r}{ 11.52 } & \multicolumn{1}{r}{ 17.81 } & \multicolumn{1}{r}{ 82.66 } & \multicolumn{1}{r}{ 11.64 } & \multicolumn{1}{r}{ 99.54 } & \multicolumn{1}{r}{ 17.43 } & \multicolumn{1}{r}{ 16.71 } & \multicolumn{1}{r}{ 2.49 }\\

\textbf{Type 2} & \multicolumn{1}{r}{ 25.61 } & \multicolumn{1}{r}{ 29.46 } & \multicolumn{1}{r}{ 0.13 } & \multicolumn{1}{r}{ 0.15 } & \multicolumn{1}{r}{ 11.94 } & \multicolumn{1}{r}{ 30.03 } & \multicolumn{1}{r}{ 2.15 } & \multicolumn{1}{r}{ 8.54 } & \multicolumn{1}{r}{ 0.05 } & \multicolumn{1}{r}{ 22.79 } & \multicolumn{1}{r}{ 4.38 } & \multicolumn{1}{r}{ 2.37 }\\ 

\textbf{Type 3} & \multicolumn{1}{r}{ 9.6 } & \multicolumn{1}{r}{ 6.43 } & \multicolumn{1}{r}{ 1.39 } & \multicolumn{1}{r}{ 2.95 } & \multicolumn{1}{r}{ 19.56 } & \multicolumn{1}{r}{ 12.41 } & \multicolumn{1}{r}{ 10.21 } & \multicolumn{1}{r}{ 23.22 } & \multicolumn{1}{r}{ 0.05 } & \multicolumn{1}{r}{ 18.43 } & \multicolumn{1}{r}{ 0.78 } & \multicolumn{1}{r}{ 0.03 }\\ 

\textbf{Type 4} & \multicolumn{1}{r}{ 27.89 } & \multicolumn{1}{r}{ 32.48 } & \multicolumn{1}{r}{ 0.1 } & \multicolumn{1}{r}{ 0.10 } & \multicolumn{1}{r}{ 6.96 } & \multicolumn{1}{r}{ 20.48 } & \multicolumn{1}{r}{ 1.39 } & \multicolumn{1}{r}{ 7.95 } & \multicolumn{1}{r}{ 0.2 } & \multicolumn{1}{r}{ 13.21 } & \multicolumn{1}{r}{ 76.17 } & \multicolumn{1}{r}{ 94.99 }\\ 

\textbf{Type 5} & \multicolumn{1}{r}{ 1.56 } & \multicolumn{1}{r}{ 1.51 } & \multicolumn{1}{r}{ 0.00 } & \multicolumn{1}{r}{0.00 } & \multicolumn{1}{r}{ 0.85 } & \multicolumn{1}{r}{ 0.33 } & \multicolumn{1}{r}{ 0.02 } & \multicolumn{1}{r}{ 0.18 } & \multicolumn{1}{r}{ 0.05 } & \multicolumn{1}{r}{ 1.68 } & \multicolumn{1}{r}{ 0.74 } & \multicolumn{1}{r}{ 0.1 }\\

\textbf{Type 6} & \multicolumn{1}{r}{ 0.41 } & \multicolumn{1}{r}{ 0.18 } & \multicolumn{1}{r}{ 0.00 } & \multicolumn{1}{r}{ 0.00 } & \multicolumn{1}{r}{ 0.94 } & \multicolumn{1}{r}{ 0.17 } & \multicolumn{1}{r}{ 0.02 } & \multicolumn{1}{r}{ 0.14 } & \multicolumn{1}{r}{ 0.00 } & \multicolumn{1}{r}{ 0.97 } & \multicolumn{1}{r}{ 0.08 } & \multicolumn{1}{r}{ 0.00 }\\

\textbf{Type 7} & \multicolumn{1}{r}{ 3.22 } & \multicolumn{1}{r}{ 0.91 } & \multicolumn{1}{r}{ 90.86 } & \multicolumn{1}{r}{ 91.98 } & \multicolumn{1}{r}{ 20.61 } & \multicolumn{1}{r}{ 3.16 } & \multicolumn{1}{r}{ 1.94 } & \multicolumn{1}{r}{ 25.9 } & \multicolumn{1}{r}{ 0.03 } & \multicolumn{1}{r}{ 5.75 } & \multicolumn{1}{r}{ 0.72 } & \multicolumn{1}{r}{ 0.00 }\\

\textbf{Type 8} & \multicolumn{1}{r}{ 0.44 } & \multicolumn{1}{r}{ 0.24 } & \multicolumn{1}{r}{ 0.00 } & \multicolumn{1}{r}{ 0.00 } & \multicolumn{1}{r}{ 1.07 } & \multicolumn{1}{r}{ 0.27 } & \multicolumn{1}{r}{ 0.01 } & \multicolumn{1}{r}{ 0.14 } & \multicolumn{1}{r}{ 0.00 } & \multicolumn{1}{r}{ 0.95 } & \multicolumn{1}{r}{ 0.04 } & \multicolumn{1}{r}{ 0.00 }\\

\textbf{Type 9} & \multicolumn{1}{r}{ 0.14 } & \multicolumn{1}{r}{ 0.16 } & \multicolumn{1}{r}{ 0.00 } & \multicolumn{1}{r}{ 0.00 } & \multicolumn{1}{r}{ 0.12 } & \multicolumn{1}{r}{ 0.06 } & \multicolumn{1}{r}{ 0.00 } & \multicolumn{1}{r}{ 0.01 } & \multicolumn{1}{r}{ 0.00 } & \multicolumn{1}{r}{ 0.19 } & \multicolumn{1}{r}{ 0.00 } & \multicolumn{1}{r}{ 0.00 }\\

\textbf{Type 10} & \multicolumn{1}{r}{ 12.94 } & \multicolumn{1}{r}{ 10.37 } & \multicolumn{1}{r}{ 1.1 } & \multicolumn{1}{r}{ 3.01 } & \multicolumn{1}{r}{ 15.98 } & \multicolumn{1}{r}{ 14.4 } & \multicolumn{1}{r}{ 1.5 } & \multicolumn{1}{r}{ 21.38 } & \multicolumn{1}{r}{ 0.06 } & \multicolumn{1}{r}{ 15.57 } & \multicolumn{1}{r}{ 0.24 } & \multicolumn{1}{r}{ 0.01 }\\

\textbf{Type 11} & \multicolumn{1}{r}{ 0.69 } & \multicolumn{1}{r}{ 0.29 } & \multicolumn{1}{r}{ 0.00 } & \multicolumn{1}{r}{ 0.01 } & \multicolumn{1}{r}{ 1.07 } & \multicolumn{1}{r}{ 0.2 } & \multicolumn{1}{r}{ 0.02 } & \multicolumn{1}{r}{ 0.15 } & \multicolumn{1}{r}{ 0.02 } & \multicolumn{1}{r}{ 0.84 } & \multicolumn{1}{r}{ 0.12 } & \multicolumn{1}{r}{ 0.00 }\\

\textbf{Type 12} & \multicolumn{1}{r}{ 1.32 } & \multicolumn{1}{r}{ 0.15 } & \multicolumn{1}{r}{ 0.16 } & \multicolumn{1}{r}{ 0.06 } & \multicolumn{1}{r}{ 4.47 } & \multicolumn{1}{r}{ 0.47 } & \multicolumn{1}{r}{ 0.03 } & \multicolumn{1}{r}{ 0.42 } & \multicolumn{1}{r}{ 0.00 } & \multicolumn{1}{r}{ 1.63 } & \multicolumn{1}{r}{ 0.02 } & \multicolumn{1}{r}{ 0.00 }\\

\textbf{Type 13} & \multicolumn{1}{r}{ 1.14 } & \multicolumn{1}{r}{ 0.04 } & \multicolumn{1}{r}{ 6.1 } & \multicolumn{1}{r}{ 1.17 } & \multicolumn{1}{r}{ 4.92 } & \multicolumn{1}{r}{ 0.21 } & \multicolumn{1}{r}{ 0.04 } & \multicolumn{1}{r}{ 0.32 } & \multicolumn{1}{r}{ 0.00 } & \multicolumn{1}{r}{ 0.56 } & \multicolumn{1}{r}{ 0.00 } & \multicolumn{1}{r}{ 0.00 }\\ \hline
\end{tabular}
\end{table*}

\begin{table}[h!]
\caption{Network diameter for 100 random network versions.} 
\label{A1_Table}
\begin{tabular}{lC{3cm}C{3cm}}
\hline
\multirow{2}{*}{} & \multicolumn{2}{c}{\textbf{mean (stdev) diameter}} \\ \cline{2-3} 
 & Bot & Human \\ \hline
\textbf{Party} & 45.25 (5.85) & 5 (0) \\
\textbf{Friendship} & 28.70 (3.85) & 10.10 (0.33) \\
\textbf{Trade} & 22.07 (1.22) & 12.87 (0.57) \\ 
\textbf{Whisper} & 29.92 (4.41) & 6 (0) \\ 
\textbf{Mail} & 20.46 (1.19) & 24.33 (1.17) \\
\textbf{Shop} & 24.57 (4.97) & 39.47 (2.62) \\ \hline
\end{tabular}
\end{table}

\begin{table}[h!]
\caption{Diameter comparison for bot and human users.}\label{A3_Table}
\begin{tabular}{lC{3cm}C{3cm}}
\hline
\multirow{2}{*}{} & \multicolumn{2}{c}{\textbf{mean (stdev) diameter}} \\ \cline{2-3} 
 & Bot & Human \\ \hline
\textbf{Party} & 45.25 (5.85) & 5 (0) \\
\textbf{Friendship} & 28.70 (3.85) & 10.10 (0.33) \\
\textbf{Trade} & 22.07 (1.22) & 12.87 (0.57) \\ 
\textbf{Whisper} & 29.92 (4.41) & 6 (0) \\ 
\textbf{Mail} & 20.46 (1.19) & 24.33 (1.17) \\
\textbf{Shop} & 24.57 (4.97) & 39.47 (2.62) \\ \hline
\end{tabular}
\end{table}

\end{document}